\documentclass[
 aip,
 jcp,
amsfonts,
amsmath,
amssymb,
reprint,%
]{revtex4-1}

\usepackage{graphicx}
\usepackage{dcolumn}
\usepackage{bm}

\usepackage[utf8]{inputenc}
\usepackage[T1]{fontenc}
\usepackage{mathptmx, mathtools, amsmath, bm, esint, booktabs}
\usepackage{etoolbox}
\usepackage{multirow}

\usepackage[colorlinks,linkcolor=blue,urlcolor=blue,citecolor=blue]{hyperref}
\usepackage{subfig}

\def\bx{{\boldsymbol{x}}}
\def\bk{{\boldsymbol{k}}}
\def\bR{{\boldsymbol{R}}}
\begin{document}

\title{Real-space Hubbard-corrected density functional theory}
\author{Sayan Bhowmik}
\affiliation{College of Engineering, Georgia Institute of Technology, Atlanta, GA 30332, USA}

\author{Andrew J. Medford}
\affiliation{College of Engineering, Georgia Institute of Technology, Atlanta, GA 30332, USA}

\author{Phanish Suryanarayana}
\email[Email: ]{phanish.suryanarayana@ce.gatech.edu}
\affiliation{College of Engineering, Georgia Institute of Technology, Atlanta, GA 30332, USA}
\affiliation{College of Computing, Georgia Institute of Technology, Atlanta, GA 30332, USA}

\begin{abstract}
We present an accurate and efficient framework for real-space Hubbard-corrected density functional theory. In particular, we obtain expressions for the energy, atomic forces, and stress tensor suitable for  real-space finite-difference discretization, and develop a large-scale parallel implementation. We verify the accuracy of the formalism through comparisons with established planewave results.  We   demonstrate that the implementation is highly efficient and scalable, outperforming established planewave codes by more than an order of magnitude in minimum time to solution, with increasing advantages as the system size and/or number of processors is increased.  We apply this framework to examine the impact of exchange-correlation inconsistency in local atomic orbital generation and introduce a scheme for optimizing the Hubbard parameter based on hybrid functionals, both while studying TiO$_2$ polymorphs.
\end{abstract}

\maketitle
\section{Introduction}
Over the past few decades, Kohn-Sham density functional theory (DFT) \cite{hohenberg1964inhomogeneous, kohnSham1965} has emerged as a foundational tool in materials and chemical sciences research, due to the fundamental insights it provides and its strong predictive capabilities. Rooted in the first principles of quantum mechanics, DFT has gained widespread adoption due to its balance of simplicity, broad applicability, and favorable accuracy-to-cost ratio compared to other such ab initio methods \cite{burkePerspective2012,beckeFifty2014}. Despite being more computationally efficient than wavefunction-based approaches, Kohn-Sham DFT remains resource-intensive, which limits the size and complexity of systems that can be studied. These challenges are especially acute for ab initio molecular dynamics (AIMD) simulations, which may require tens to hundreds of thousands of Kohn-Sham solves to explore the desired properties or phenomena.

For a given system, the cost and accuracy of a Kohn-Sham DFT calculation are primarily governed by the choice of exchange-correlation functional, which accounts for many-body electron interactions. In the absence of a universal functional, a hierarchy of approximations categorized by Jacob's ladder~\cite{jacobsLadder} has been developed, where higher rungs generally provide improved accuracy at the expense of increased computational effort. The lowest three rungs of the ladder, corresponding to local and semilocal functionals, fail to adequately capture the strong on-site Coulomb interactions among localized electrons in strongly correlated materials, while hybrid functionals on the fourth rung can improve accuracy for such cases but incur a substantially higher computational cost. An attractive alternative is Hubbard-corrected DFT, commonly known as DFT+U, a formalism within the Kohn-Sham DFT framework in which a Hubbard-like correction is added to the Kohn-Sham energy functional based on local/semilocal exchange-correlation~\cite{anisimov1991, CAPELLE201391}. This typically involves an adjustable parameter but efforts have been made to determine it using linear response~\cite{pickett1998,linearResponseDftU2005}. The DFT+U formalism outperforms local/semilocal exchange-correlation functionals in predicting the properties of strongly correlated materials, while incurring significantly lower computational cost than hybrid functionals. In particular, DFT+U has been shown to improve predictions of structural properties~\cite{linearResponseDftU2005, dftu_structProp}, magnetic properties~\cite{dftu_magProp, martin2020electronic}, band gaps~\cite{martin2020electronic}, phase stability~\cite{dftu_phase}, and chemical reaction energies~\cite{hjk2006, dftu_redox, dftu_chemRxn}. The DFT+U model can be extended to include inter-site corrections, alongside the conventional on-site term, for systems in which significant orbital hybridization causes electron localization to span multiple sites, resulting in the DFT+U+V model~\cite{dft_u_v}.

The planewave method \cite{martin2020electronic} has been one of the most widely adopted strategies for solving the Kohn-Sham equations within the pseudopotential approximation \cite{kresse1996, CASTEP, ABINIT, Espresso, CPMD, DFT++, gygi2008architecture, valiev2010nwchem}. Its use of a Fourier basis ensures high accuracy and enables efficient computations on moderate resources through well-optimized Fast Fourier Transforms (FFTs). However, the method is intrinsically limited to periodic boundary conditions, and the global nature of the basis poses challenges for scalability on parallel computing platforms. These limitations have prompted the development of alternative strategies that employ systematically improvable, localized representations \cite{Becke1989, chelikowsky1994, wavelets2008, seitsonen1995, whiteFEM1989, IWATA20102339, tsuchida1995, qimenDDBP, SURYANARAYANA2010256, SURYANARAYANA20115226, onetep, CONQUEST, MOTAMARRI2020106853, octopus, rmgdft, FATTEBERT199975, shimojo2001linear, arias1999wav, pask2005femeth, lin2012adaptive, xu2021sparc}. Among these, real-space finite-difference methods are among the most mature and widely adopted, as they offer high computational locality and naturally accommodate Dirichlet, periodic, and Bloch-periodic boundary conditions. These methods often outperform their planewave counterparts for local,  semilocal, and hybrid exchange-correlation functionals, with performance benefits that grow with increasing processor counts \cite{xu2021sparc, zhang2024sparc, XinHybrid2024}. Furthermore, they have demonstrated scalability to systems approaching and exceeding a million electrons~\cite{gavini2022roadmap, fattebert2016modeling, mehmet2023}.  However, although the DFT+U formalism has been implemented in real-space codes~\cite{GPAW, octopus, rmgdft, MOTAMARRI2020106853}, some of which have been discussed in the literature~\cite{GPAW, octopus_hub1, octopus_hub2}, its accuracy and efficiency within the real-space framework, particularly in the context of parallel computations, have not been demonstrated heretofore. Furthermore, to the best of our knowledge, an implementation employing exchange–correlation-consistent local orbitals has not been developed heretofore.

In this work, we develop an accurate and efficient real-space framework for DFT+U, including expressions for the energy, atomic forces, and stress tensor that are amenable to the finite-difference discretization, along with a scalable parallel implementation. We validate the framework  against established planewave results and show that it can achieve over an order-of-magnitude speedup in minimum time to solution, with increasing benefits at larger number of processors and/or system size. We apply the framework to study TiO$_2$ polymorphs, examining the role of exchange-correlation inconsistency in local atomic orbital generation and introducing a hybrid-functional-based scheme for  Hubbard parameter optimization.

The remainder of this paper is organized as follows. In Section~\ref{Sec:Real_space_formulation}, we present the real-space formulation and implementation for DFT+U. Next, we verify its accuracy and efficiency, and apply it to study TiO$_2$ polymorphs in Section~\ref{Sec:Results}. Finally, we provide concluding remarks in Section~\ref{Sec:Concluding_remarks}.


\section{Real-space DFT+U}\label{Sec:Real_space_formulation}

In the DFT+U formalism, a Hubbard-like term is included in the energy functional: 
\begin{align}
    E_{\text{DFT+U}} &= E_{\text{DFT}} + E_U \,, \label{eq:E_full_1}
\end{align}  
where \( E_{\text{DFT}} \) is the energy functional corresponding to local/semilocal exchange-correlation in Kohn-Sham DFT, and \( E_U \) is the Hubbard correction, comprised of a model interaction term and a double-counting term that compensates for the portion of electron–electron interactions already included in $E_{\text{DFT}}$.

We now discuss the DFT+U formalism by Dudarev \textit{et al}. \cite{dudarev1998} in the context of the real-space method. In particular, we discuss its formulation and implementation within the SPARC electronic structure code \cite{xu2021sparc, zhang2024sparc}  in Sections~\ref{Subsec:Formulation} and \ref{Subsec:Implementation}, respectively. 

\subsection{Formulation} \label{Subsec:Formulation}

In Section~\ref{Subsec:Formulation}, we present the Hubbard energy correction (Section~\ref{SubSubsec:Energy}), followed by the corresponding potential (Section~\ref{SubSubsec:Potential}), Hellmann-Feynman atomic forces (Section~\ref{SubSubsec:Forces}), and Hellmann-Feynman stress tensor (Section~\ref{SubSubsec:Stress}). The expressions for the energy and potential presented here are well established and, aside from being discretized in real and Fourier space, respectively, are common to both planewave and real-space formulations. In contrast, the expressions for the force and stress tensor presented here are specific to the real-space finite-difference method. Indeed, in the planewave approach, high-frequency components of quantities are truncated above a prescribed cutoff, thereby preserving translational symmetry and eliminating the eggbox effect. In the real-space method, however,  eggbox effects can be significant. Therefore, beyond achieving an efficient implementation, the primary challenge in the real-space method lies in maintaining high accuracy without resorting to excessively fine grids.
\subsubsection{Energy}\label{SubSubsec:Energy}
The Hubbard energy in the rotationally invariant scheme proposed by Dudarev \textit{et al} \cite{dudarev1998}  takes the form:
\begin{align}
    E_U = \sum_{\sigma I \ell} \frac{\widetilde{U}_{I \ell}^{\sigma}}{2} \text{Tr} \left\{ \mathcal{D}_{I \ell}^{\sigma} - (\mathcal{D}_{I \ell}^{\sigma} )^2 \right\}  \,, \label{eq:E_U}
\end{align}
$\sigma \in \{ \uparrow, \downarrow \}$ denotes the spin, $I$ is an index that runs over all the atoms in the cell $\Omega$, $\ell$ is the angular quantum number, $\rm Tr \{ .\}$ denotes the trace, $\widetilde{U}$ is the effective Hubbard interaction parameter, and $\mathcal{D}$ is the local orbital occupation (Hermitian) matrix of size $2 \ell + 1$ with components:
\begin{align}\label{eq:orbital_occup_ext}
   ( \mathcal{D}_{I \ell}^{\sigma} )_{m m'}  = & \sum_{j} \fint_{BZ} f^\sigma_j(\bk) \left( \int_{\Omega}\psi^{\sigma*}_{j}(\bx,\bk)\varphi_{I\ell m'}(\bx,\bk, \bR_I) d\bx \right) \nonumber \\
    &\times \left(\int_{\Omega} \varphi^{*}_{I \ell m}(\bx,\bk,\bR_I)\psi^\sigma_{j}(\bx,\bk) d\bx \right) d\bk \,.
\end{align}
Above, the indices $m$ and $m'$ signify the magnetic quantum number, the index $n$ runs over all the Kohn-Sham states, $\fint_{BZ}$ denotes the average over the Brillouin zone, $(.)^*$ represents the complex conjugate, $\bk$ denotes a wavevector in the Brillouin zone, $\bx$ represents a spatial point in $\Omega$,  $\psi$ are the Kohn-Sham orbitals with occupations $f$, and $\varphi$ are Bloch periodically mapped atomic orbitals, i.e.,  
\begin{align}
\varphi_{I\ell m}(\bx,\bk,\bR_I) = \sum_{\tilde{I}} \hat{\varphi}_{\tilde{I}\ell m}(|\bx-\bR_I|) e^{-\mathrm{i}\bk\cdot(\bR_I - \bR_{\tilde{I}})} \,,
\end{align}
where the index $\tilde{I}$ runs over the $I$-th atom and its periodic images, $\mathrm{i} = \sqrt{-1}$, and $\hat{\varphi}$ are the atom-centered atomic orbitals.

\subsubsection{Potential}\label{SubSubsec:Potential}
The Hubbard potential $V_U^\sigma$,  i.e., potential operator corresponding to the energy in Eq.~\ref{eq:E_U} that enters the Hamiltonian for a given spin channel, can be obtained from the variational/functional derivative of the Hubbard  energy  with respect to the complex conjugate of the orbitals:
\begin{align}
\frac{\delta E_U}{\delta \psi_n^{\sigma *}} = f_n^{\sigma} V_U \psi_n^{\sigma} \,.
\end{align}
Using the chain rule, the functional derivative can be written as:
\begin{align}
\frac{\delta E_U}{\delta \psi_n^{\sigma *}} = \sum_{\sigma' I' \ell' m m'} 
\frac{\partial E_{U}}{\partial ( \mathcal{D}_{I' \ell '}^{\sigma '} )_{m m'}} \frac{\delta ( \mathcal{D}_{I' \ell '}^{\sigma '} )_{m m'}}{\delta \psi_n^{\sigma*}}  \,. 
\end{align}
It can be shown that:
\begin{align}
\frac{\partial E_{U}}{\partial ( \mathcal{D}_{I' \ell '}^{\sigma '} )_{m m'}}   & =   \frac{\widetilde{U}_{I' \ell '}^{\sigma '}}{2} \left( \delta_{m m'} -  2 ( \mathcal{D}_{I' \ell '}^{\sigma '} )_{m m'} \right)\,, \label{Eq:DerED}\\
\frac{\delta ( \mathcal{D}_{I' \ell '}^{\sigma '} )_{m m'}}{\delta \psi_n^{\sigma*}}  & = \delta_{\sigma \sigma'} f_n^{\sigma '}(\bk) \varphi_{I' \ell ' m'}(\bx,\bk,\bR_{I'}) \nonumber \\ 
 & \times  \left( \int_{\Omega}\varphi_{I' \ell ' m}^*(\bx,\bk,\bR_{I'})  \psi^{\sigma '}_{n}(\bx,\bk)  d\bx \right) \,,
\end{align}
where $\delta_{(.)}$ denotes the Kronecker delta. It therefore follows that the Hubbard potential when operated on the orbitals takes the form:
\begin{align}
V_{U}^{\sigma} & \psi^{\sigma}_n(\bx,\bk)  =  \sum_{I \ell m m'}  \frac{\widetilde{U}_{I \ell}^{\sigma}}{2} \left( \delta_{m m'} -  2 ( \mathcal{D}_{I \ell}^{\sigma} \right)_{m m'} ) \nonumber \\
& \times \varphi_{I \ell m'}(\bx,\bk)  \left(\int_{\Omega}\varphi_{I\ell m}^*(\bx,\bk)\psi^\sigma_{n}(\bx,\bk) \, d\bx \right)  \,.
\end{align}
Indeed, unlike the potential arising from local/semilocal exchange-correlation functionals, the Hubbard potential is orbital dependent.  
\subsubsection{Atomic forces}\label{SubSubsec:Forces}
The Hellmann-Feynman atomic forces can be calculated once the electronic ground state has been determined. The Hubbard atomic forces on the $J$-th atom can be defined as:
\begin{align}\label{eq:Force1}
    \boldsymbol{f}_U^J = -\frac{\partial E_U}{\partial \bR_J} \,.
\end{align}
Using the chain rule, the Hubbard forces can be written as:
\begin{align}
\boldsymbol{f}_U^J  = - \sum_{\sigma' I' \ell' m m'} 
\frac{\partial E_{U}}{\partial ( \mathcal{D}_{I' \ell '}^{\sigma '} )_{m m'}} \frac{\partial ( \mathcal{D}_{I' \ell '}^{\sigma '} )_{m m'}}{\partial \bR_J}  \,.
\end{align}
The first derivative is given by Eq.~\ref{Eq:DerED}, and for the second derivative, it can be shown that:
\begin{widetext}
\begin{align}
 \frac{\partial ( \mathcal{D}_{I' \ell '}^{\sigma '} )_{m m'}}{\partial \bR_J} =  
2\sum_j \fint_{BZ} f_j^{\sigma'}(\bk)  \Re\Bigg\{ \left( \int_{\Omega}{\nabla \psi_j^{\sigma'*}(\bx,\bk)}{\varphi_{J \ell ' m'}(\bx,\bk,\bR_J) d\bx}  \right)
\left(\int_{\Omega}{\varphi^{*}_{J \ell ' m}(\bx,\bk,\bR_J)}{\psi^{\sigma '}_{j}(\bx,\bk) d\bx} \right) \Bigg\} d\bk \,, \label{Eq:DerR}
\end{align}
\end{widetext}
where $\Re \{.\}$ denotes the real part of the complex-valued quantity. In arriving at the expression,  we have transferred the derivatives from the atomic orbitals to the Kohn-Sham orbitals, as done for the nonlocal component of the forces \cite{hirose2005first, ghosh2017sparc2} and the Hubbard atomic forces \cite{octopus_hub1}. This allows the use of the finite-difference approximation to the gradient, which is readily available. In so doing, we have found that the quality of the forces is significantly improved, since the   Kohn-Sham orbitals are generally smoother than the atomic orbitals. Moreover, the derivatives of the Kohn-Sham orbitals are available as part of the nonlocal pseudopotential force calculation in SPARC. The expression for the atomic forces then takes the form:
\begin{widetext}
\begin{align}
\boldsymbol{f}_U^J =  \sum_{\sigma \ell m m' j}  \widetilde{U}_{J \ell}^{\sigma} \left(  2 ( \mathcal{D}_{J \ell}^{\sigma} \right)_{m m'} - \delta_{m m'})  \fint_{BZ} f_j^\sigma(\bk)  \Re\Bigg\{ \left(\int_{\Omega}{\nabla \psi_j^{\sigma*}(\bx,\bk)}{\varphi_{J \ell m'}(\bx,\bk,\bR_J) d\bx} \right) \left( \int_{\Omega}{\varphi^{*}_{J \ell m}(\bx,\bk,\bR_J)}{\psi^\sigma_{j}(\bx,\bk) d\bx} \right) \Bigg\} d\bk \,.
\end{align}
\end{widetext}
Note that the above expression for the Hubbard atomic forces has been specifically formulated for the real-space finite-difference method. Indeed, its form can vary considerably depending on the real-space discretization employed, as exemplified by the nonlocal pseudopotential force expression in the finite-element method~\cite{Motamarri_force_2018}. 

\subsubsection{Stress tensor}\label{SubSubsec:Stress}
The Hellmann-Feynman stress tensor can also be calculated once the electronic ground state has been determined. The  Hubbard stress tensor components can be defined as \cite{sharma2018stress}:
\begin{align}\label{Eq:StressDef}
(\boldsymbol{\sigma}_U)_{\alpha \beta} &= \frac{1}{|\Omega|} \frac{\partial E_{\text{U}}}{\partial (\boldsymbol{F})_{\alpha\beta}}\Bigg{|}_\mathcal{G} \,, \quad \alpha, \beta \in \{1, 2, 3\} \,,
\end{align}
where $|\Omega|$ denotes the measure of the cell $\Omega$, e.g., volume of the cell when the system is extended/periodic in all three directions, $\boldsymbol{F}$ is the deformation gradient, and $\mathcal{G}$ represents the electronic ground state corresponding to the undeformed configuration, i.e., $\bm{F} = \bm{I}$, with $\bm{I}$ being the identity matrix. Using the chain rule, the  Hubbard stress tensor components can be written as:
\begin{align}
(\boldsymbol{\sigma}_U)_{\alpha \beta} = \sum_{\sigma' I' \ell' m m'} 
\frac{\partial E_{U}}{\partial ( \mathcal{D}_{I' \ell '}^{\sigma '} )_{m m'}} \frac{\delta ( \mathcal{D}_{I ' \ell '}^{\sigma '} )_{m m'}}{\delta (\boldsymbol{F})_{\alpha\beta}} \Bigg{|}_\mathcal{G} \,.
\end{align}
The first derivative is given by Eq.~\ref{Eq:DerED}, evaluated at the electronic ground state, and for the second derivative, it can be shown that:
\begin{widetext}
\begin{align}
\frac{\delta ( \mathcal{D}_{I' \ell '}^{\sigma '} )_{m m'}}{\delta (\boldsymbol{F})_{\alpha\beta}} \Bigg{|}_\mathcal{G}  = - 2 \sum_{j}  \fint_{BZ}  f_j^{\sigma'}(\bk)  \Re & \Bigg\{ \left( \int_{\Omega} \psi^{{\sigma'}*}_j (\bx,\bk) \varphi_{I \ell ' m'}(\bx,\bk, \bm{R}_{I'}) d\bx \right) \nonumber \\
&  \times
\left( \sum_{\tilde{I}} \int_{\Omega} \varphi^{*}_{\tilde{I} \ell ' m} e^{i \bk \cdot \left( \bm{R}_{I'} - \bm{R}_{\tilde{I}}\right) }\left( \bm{S}^{T}\bx - \bm{S}^{T}\bm{R}_{I'} \right)_{\beta} \nabla_{\alpha} \psi_{j}^{{\sigma'}}(\bx,\bk) d\bx \right) \Bigg\}  d\bk  
 - \delta_{\alpha \beta} ( \mathcal{D}_{I' \ell '}^{\sigma '} )_{m m'} \,,
\end{align}
\end{widetext}
where $(.)_{\beta}$ refers to the $\beta$-component of the vector, $\nabla_\alpha$ refers to the $\alpha$-component of the gradient vector in the Cartesian coordinate system, and $\bm{S}$ is the transformation matrix that relates the Cartesian unit vectors with the lattice unit vectors, with $\bm{S} = \bm{I}$ for an orthogonal/orthorhombic cell. In arriving at the above expression,  we have followed the derivation presented in previous work for the nonlocal pseudopotential stress tensor in real-space DFT \citep{sharma2018stress}. In particular, we have again transferred the derivatives from the atomic orbitals to the Kohn-Sham orbitals. This allows the use of the finite-difference approximation to the gradient, which is readily available. Such a strategy provides advantages similar to those discussed above for the forces, namely, significantly higher quality of stresses and availability of the derivative of the Kohn-Sham orbitals for the calculation of the nonlocal pseudopotential stresses in SPARC.  The expression for the Hubbard stress tensor components then takes the form:
\begin{widetext}
\begin{align}
(\boldsymbol{\sigma}_U)_{\alpha \beta} = 
\sum_{\sigma I \ell m m'} 
\frac{\widetilde{U}_{I\ell}^{\sigma}}{2} & \left( \delta_{m m'} - 2 ( \mathcal{D}_{I\ell}^{\sigma} )_{m m'} \right)  \Bigg( - 2 \sum_{j}  \fint_{BZ}  f_j^{\sigma'}(\bk)  \Re \Bigg\{ \int_{\Omega} \left( \psi^{{\sigma'}*}_j (\bx,\bk) \varphi_{I \ell ' m'}(\bx,\bk, \bm{R}_{I'}) d\bx \right) \nonumber \\
&  \times
\left(\sum_{\tilde{I}} \int_{\Omega} \varphi^{*}_{\tilde{I} \ell ' m} e^{i \bk \cdot \left( \bm{R}_{I'} - \bm{R}_{\tilde{I}}\right) }\left( \bm{S}^{T}\bx - \bm{S}^{T}\bm{R}_{I'} \right)_{\beta} \nabla_{\alpha} \psi_{j}^{{\sigma'}}(\bx,\bk) d\bx \right)  \Bigg\}  d\bk  
 - \delta_{\alpha \beta} ( \mathcal{D}_{I' \ell '}^{\sigma '} )_{m m'} \Bigg) \,.
\end{align}
\end{widetext}
Note that the above expression for the Hubbard stress tensor components has been specifically formulated for the real-space finite-difference method, and has not been presented in the literature heretofore. Indeed, its form can vary considerably depending on the real-space discretization employed, as exemplified by the nonlocal pseudopotential stress expression in the finite-element method~\cite{Motamarri_force_2018}. 

\subsection{Implementation} \label{Subsec:Implementation}
We have implemented the real-space DFT+U formalism described above in the SPARC electronic structure code \cite{xu2021sparc,zhang2024sparc}. In particular, we perform a uniform real-space discretization of the equations, using high-order centered finite-differences for approximating derivatives and the trapezoidal rule for integrations. Specifically, consider a cell $\Omega$ with side lengths $L_1$, $L_2$, and $L_3$, discretized using a uniform finite-difference grid with spacings $h_1$, $h_2$, and $h_3$, respectively, such that $L_1 = n_1 h_1$, $L_2 = n_2 h_2$, and $L_3 = n_3 h_3$, where $n_1, n_2, n_3 \in \mathbb{N}$. For each node in the finite-difference grid, indexed by $(i,j,k)$ with $i = 1, 2, \ldots, n_1$, $j = 1, 2, \ldots, n_2$, and $k = 1, 2, \ldots, n_3$, the gradient of any function $f$ (orbitals in the present case) is discretized as:\cite{ghosh2017sparcI,ghosh2017sparc2}:
\begin{align}
    \nabla f \Big|^{(i,j,k)} \approx \sum_{p=1}^{n_0}\Bigg[& \widetilde{w}_{p,1} \Big(f^{(i+p,j,k)} - f^{(i-p,j,k)} \Big) \bm{\hat{e}_1} \nonumber \\
    &+ \widetilde{w}_{p,2} \Big(f^{(i,j+p,k)} - f^{(i,j-p,k)} \Big) \bm{\hat{e}_2} \nonumber \\
    &+ \widetilde{w}_{p,3} \Big(f^{(i,j,k+p)} - f^{(i,j,k-p)} \Big) \bm{\hat{e}_3} \Bigg] \,,
\end{align}
where $2n_0$ is the order of finite-difference approximation, $\bm{\hat{e}_1},\ \bm{\hat{e}_2}$ and $\bm{\hat{e}_3}$ are unit vectors along the edges of $\Omega$, and the stencil weights
\begin{align}
    \widetilde{w}_{p,a} = \frac{(-1)^{p+1}}{h_a p} \frac{(n_0 !)^2}{(n_0 - p)! (n_0+p)!},\ p=1,2,\cdots,n_0 \,.
\end{align}
Any index that is not in the grid is periodically mapped back with the appropriate Bloch factor. The spatial integrations are evaluated as:
\begin{align}
    \int_{\Omega} f(\bx) d\bx = h_1 h_2 h_3 \det{(S)} \sum_{i=1}^{n_1} \sum_{j=2}^{n_2} \sum_{k=1}^{n_3} f^{(i,j,k)} \,,
\end{align}
where $\det{(.)}$ represents the determinant of the matrix.

The atomic orbitals are constructed on the real-space grid by combining real spherical harmonics with exchange-correlation-consistent radial atomic orbitals, obtained from a recently developed solver that utilizes a spectral scheme based on Chebyshev polynomials for atomic structure calculations \cite{bhowmik2025}. This enables the use of exchange--correlation consistent atomic orbitals  for higher-rung exchange--correlation functionals, a feature that is not generally available in other DFT codes, as they typically rely on atomic orbitals provided within the pseudopotential file. The spatial extent of the radial atomic orbitals is determined by a threshold on their value; components falling below this threshold are truncated to zero and therefore not stored.

Given the similarity between the atomic orbitals and the nonlocal pseudopotential projectors, both being atom-centered and spatially localized, the implementation of DFT+U in SPARC closely mirrors that of the nonlocal pseudopotential. In particular, for a given spin and Brillouin zone wavevector, the stored atomic orbitals are Bloch periodically mapped, after which all integrals associated with a given atom are computed simultaneously. This process is then repeated for each atom to which the Hubbard correction is applied. Since the implementation employs Bloch boundary conditions for the orbitals, rather than the Bloch ansatz used in most other real-space DFT codes, the number of operations associated with DFT+U computations in SPARC is substantially reduced, particularly as the system size increases.

The electronic ground state is determined using  Chebyshev filtered subspace iteration (CheFSI)\cite{ChefsiI}, which is accelerated using the real-space preconditioned \cite{KUMAR2020136983} Pulay mixing \cite{pulay1980convergence}. The guess for the local orbital occupation matrix corresponding to an atom and angular momentum is set to be a diagonal matrix, with each entry  being the average electron occupation of that  state in the isolated atom. Since the Hamiltonian depends on the orbitals through the local orbital occupation matrix, the matrix needs to be updated during each self-consistent field (SCF) iteration. To accelerate convergence, the mixing coefficients used for the electron density are also applied to the local orbital occupation matrices. As an alternative, one could use an outer loop to update the occupation matrix, holding it fixed during the inner SCF loop; however, we have found the current strategy to be more efficient.  In AIMD simulations, similar to the electron density \cite{ALFE1999_chargeExtrap}, the occupation matrix is split into a sum of atomic initial guess and a residual difference, with the occupation matrix at the next time step obtained by combining the atomic initial guess with a second-order extrapolation of the residual. We have developed a four-level parallelization hierarchy, distributing processors across spin, Bloch wavevectors, bands, and the spatial domain, in that order.  Although these different levels of parallelization are generally available in mature DFT codes, both their ordering and the relative emphasis placed on each, particularly between the domain and band levels, can vary, especially in real-space implementations. These choices can significantly affect communication patterns in the DFT+U calculations and, therefore the computational efficiency for a given number of processors.

The implementation of DFT+U in SPARC has been released (\url{https://github.com/SPARC-X/SPARC}). It has also been released in the M-SPARC electronic structure code (\url{https://github.com/SPARC-X/M-SPARC}) \cite{xu2020m}, which is a Matlab version of SPARC that can be used for rapid prototyping. It is worth noting that though the discussion in this work has focused on extended systems, the formulation and implementation are compatible with Dirichlet and periodic boundary conditions, and combinations thererof.

\section{Results and Discussion}\label{Sec:Results}
We now assess the accuracy and performance of the developed real-space DFT+U framework in  Sections~\ref{Subsec:AccConv} and~\ref{Subsec:Performance}, respectively, and subsequently apply it to the study of TiO$_2$ polymorphs in Section~\ref{Subsec:Polymorph}. All the data can be found in the Supplementary Material.

\subsection{Accuracy and convergence} \label{Subsec:AccConv}
We first assess the accuracy and convergence of the real-space DFT+U framework. We consider four systems: 7-atom non-orthogonal cell of Br$_4$FeLi$_2$, 7-atom non-orthogonal cell of FeLi$_4$N$_2$, 16-atom orthogonal cell of MoO$_3$, and 6-atom orthogonal cell of rutile TiO$_2$, which represent a diverse set of transition-metal systems. To ensure numerically meaningful forces and stresses, the atom positions and cell lengths are perturbed from their equilibrium values. In particular, the atom positions and cell dimensions are chosen such that the maximum atomic force and stress exceeds $0.01$~ha/bohr and $1$ GPa, respectively. The exact atomic configurations considered, along with the corresponding atomic forces and stresses, can be found in the Supplementary Material. We perform Brillouin zone integration using an uniform wavevector grid with spacing of $\sim 0.22 \,\, \text{bohr}^{-1}$, and  spin-unpolarized calculations are employed in all cases except FeLi$_4$N$_2$. All simulations use the  Perdew-Burke-Ernzerhof (PBE) exchange-correlation functional~\cite{perdew1996generalized} with a Hubbard parameter of $U = 5$~eV for the transition metal's $d$-electrons, optimized norm-conserving Vanderbilt (ONCV) pseudopotentials \cite{hamann2013optimized} with nonlinear core corrections from the SPMS set \cite{shojaei2023soft}, and the 12-th order accurate centered finite-difference approximation.  

\begin{figure*}[htbp]
  \centering

  \subfloat[\label{fig:Energy_rcut_conv}Energy]{%
    \includegraphics[width=0.32\textwidth]{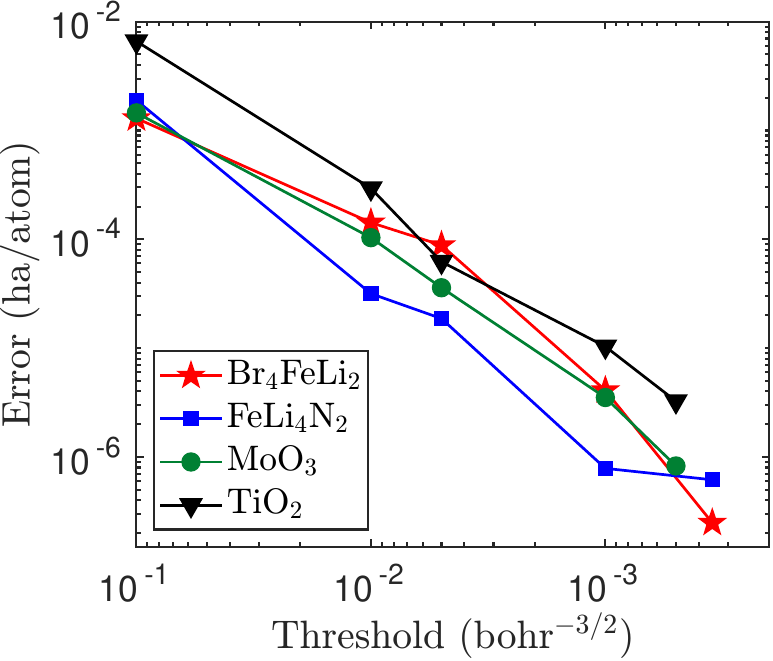}
  }
  \subfloat[\label{fig:Force_rcut_conv}Force]{%
    \includegraphics[width=0.32\textwidth]{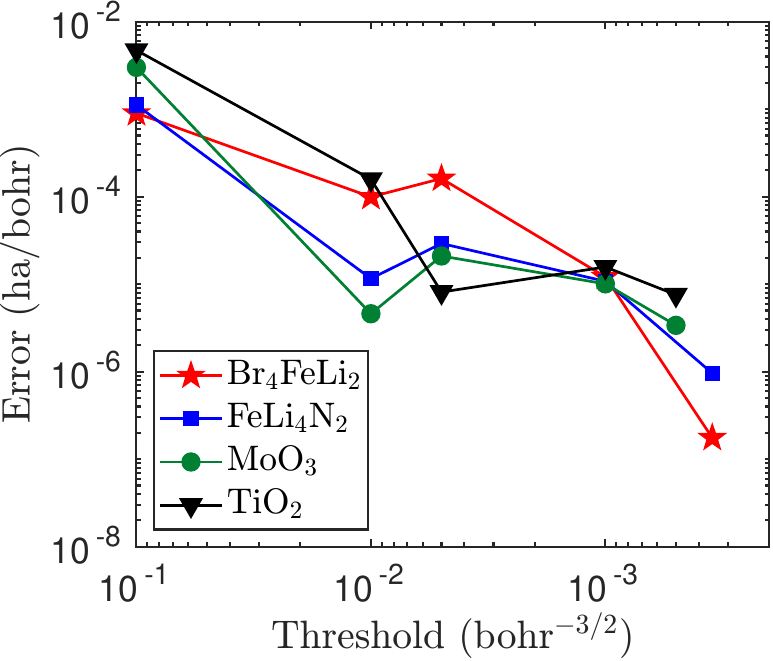}
  }
  \subfloat[\label{fig:Stress_rcut_conv}Stress]{%
    \includegraphics[width=0.32\textwidth]{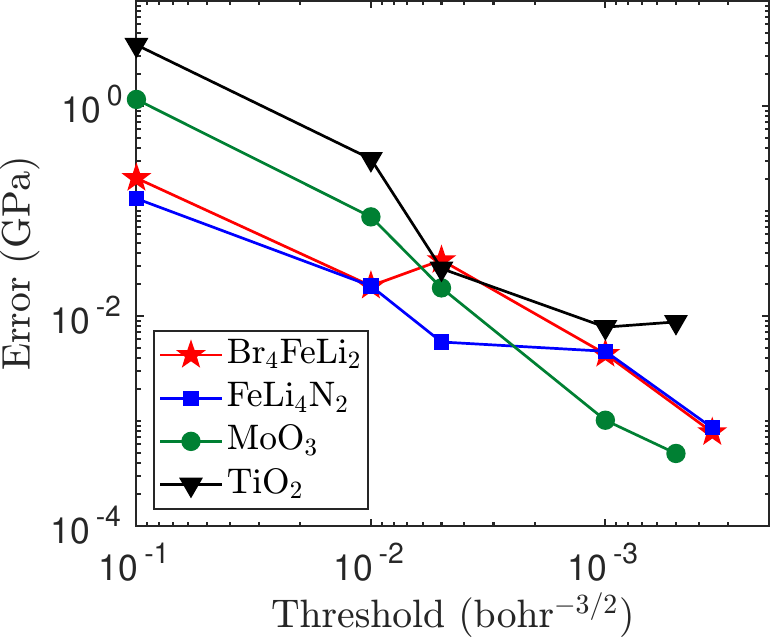}
  }
  \caption{Convergence of the energy, atomic forces, and stress tensor with respect to the local atomic orbital truncation threshold. The energy error is defined as the absolute difference, while the force and stress errors are defined as the maximum difference in any component. Results obtained with a truncation threshold of $10^{-4}$ bohr$^{-3/2}$ serve as the reference.}
  \label{Fig:rcut_conv}
\end{figure*}

In Fig.~\ref{Fig:rcut_conv}, we plot the convergence of the energy, Hellmann-Feynman atomic forces, and Hellmann-Feynman stress tensor with respect to the truncation threshold used for  the  radial atomic orbitals, using the results obtained with a  threshold of $10^{-4}$ bohr$^{-3/2}$ as reference. We use a real-space grid spacing of $\sim 0.19$~bohr for all systems, chosen to be sufficiently small to ensure physically meaningful results. We observe rapid convergence in the energy, forces, and stresses across all systems, with a threshold value of $\sim 5 \times 10^{-3}$ bohr$^{-3/2}$ sufficient to achieve an accuracy of $\sim 10^{-4}$~Ha/atom in the energy, $\sim 10^{-4}$~Ha/Bohr in the forces, and $\sim 0.01$~GPa in the stresses, which are generally well within the accuracy targets of Kohn-Sham DFT calculations. This threshold corresponds to radial distances of 6--8~bohr, depending on the atomic species, with the atomic orbitals demonstrating exponential decay  with respect to the radial distance.

\begin{figure*}[htbp]
  \centering

  \subfloat[\label{fig:Energy_conv}Energy]{%
    \includegraphics[width=0.32\textwidth]{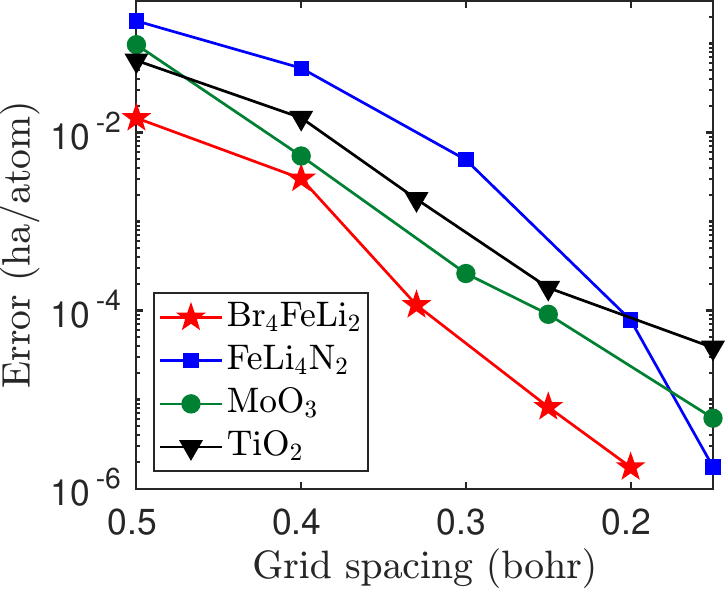}
  }
  \subfloat[\label{fig:Force_conv}Force]{%
    \includegraphics[width=0.32\textwidth]{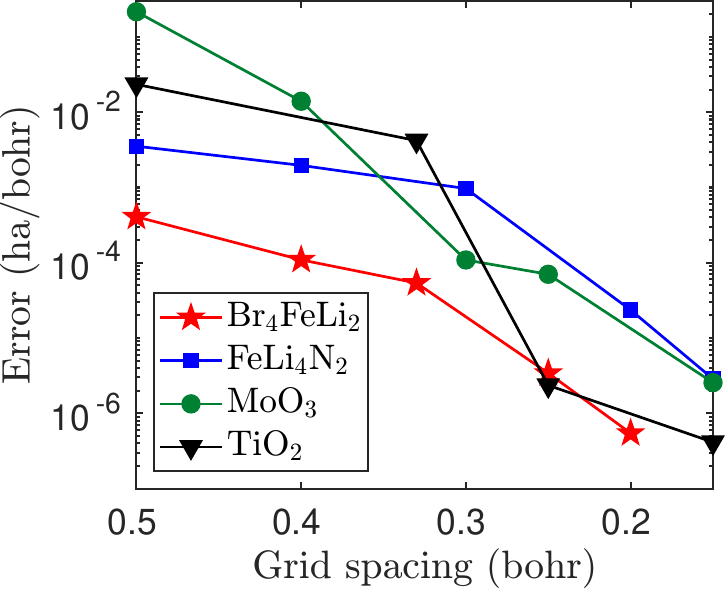}
  }
  \subfloat[\label{fig:Stress_conv}Stress]{%
    \includegraphics[width=0.32\textwidth]{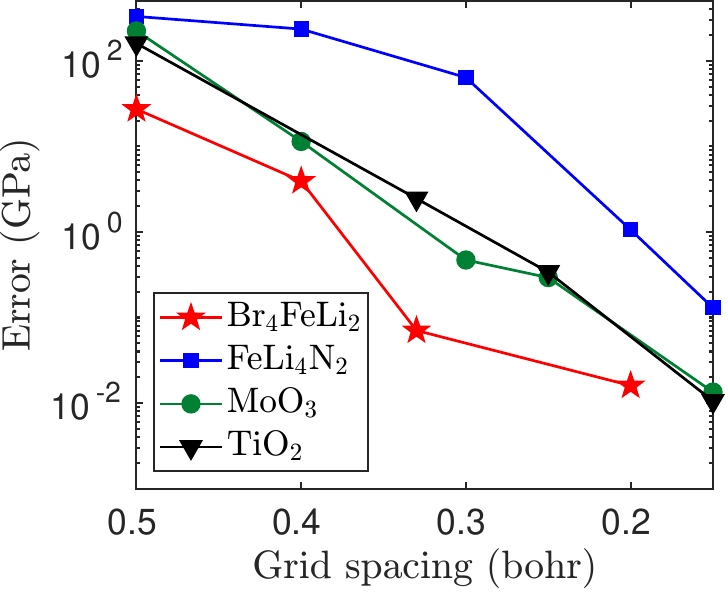}
  }
  \caption{Convergence of the energy, atomic forces, and stress tensor with respect to grid spacing. The energy error is defined as the absolute difference, while the force and stress errors are defined as the maximum difference in any component. Results obtained with a grid spacing of $\sim 0.10$~bohr serve as the reference.}
  \label{Fig:Mesh_conv}
\end{figure*}

In Fig.~\ref{Fig:Mesh_conv}, we plot the convergence of the energy, atomic forces, and stress tensor with respect to the real-space grid spacing, using the results for the grid spacing of $\sim 0.10$ bohr as reference. We choose the atomic orbital threshold to be $5 \times 10^{-3}$ bohr$^{-3/2}$ for all systems. We observe that the energy, forces, and stresses all converge systematically and rapidly --- average convergence rates obtained for the energy, forces, and stresses are $\sim 9$, $\sim 8$, and $\sim 7$, respectively, which are comparable to those obtained for local/semilocal exchange-correlation functionals \cite{ghosh2017sparc2, sharma2018stress} --- with chemical accuracy readily obtained. In particular, grid spacings of $0.33$, $0.15$, $0.20$, and $0.20$~bohr for Br$_4$FeLi$_2$, FeLi$_4$N$_2$, MoO$_3$, and TiO$_2$, respectively, are sufficient to achieve accuracies of $10^{-4}$~ha/atom in the energy, $10^{-4}$~ha/bohr in the forces, and $0.10$~GPa in the stresses.  Indeed, to achieve a given numerical accuracy, the grid spacings required for DFT+U are very similar to those needed for local/semilocal exchange-correlation, e.g., $0.20$ bohr for TiO$_2$ with PBE exchange-correlation functional yields the aforementioed accuracies, indicating  that the Hubbard correction does not necessitate a finer grid,  confirming the accuracy and efficiency of the developed framework.

In Table \ref{table:QE_comp}, we compare the SPARC results with the established planewave code Quantum Espresso (QE) \cite{Espresso}. QE reads the local atomic orbitals from the pseudopotential file (\texttt{upf} format), which are in excellent agreement with those generated by SPARC's spectral atomic solver. \cite{bhowmik2025}  In SPARC, we employ grid spacings of $0.25$, $0.15$, $0.20$, and $0.19$ for the Br$_4$FeLi$_2$, FeLi$_4$N$_2$, MoO$_3$, and TiO$_2$ systems, respectively. In addition, we employ the atomic orbital threshold to be $10^{-4}$ bohr$^{-3/2}$ for all systems. The SPARC results so obtained  are converged to within $10^{-5}$ ha/atom, $10^{-5}$ ha/bohr, and $0.1$ GPa in the energy, forces, and stresses, respectively.  In QE, we employ planewave cutoffs of $30$, $40$, $45$, and $50$ ha for the Br$_4$FeLi$_2$, FeLi$_4$N$_2$, MoO$_3$, and TiO$_2$ systems, respectively. The QE results so obtained are also converged to within $10^{-5}$ ha/atom, $10^{-5}$ ha/bohr, and $0.1$ GPa in the energy, forces, and stresses, respectively.  Indeed, the discretizations in both SPARC and QE have been chosen to achieve the target accuracy of $10^{-5}$ ha/atom, $10^{-5}$ ha/bohr, and $0.1$ GPa in the energy, forces, and stresses, respectively. We observe from the results that there is very good agreement between SPARC and QE, with the maximum difference in the energy, forces, and stresses being $8.77\times10^{-5}$ ha/atom, $1.02\times10^{-4}$ ha/bohr, and $0.17$ GPa, respectively.  These results further verify the accuracy of the developed real-space DFT+U framework.

\begin{table}[htbp]
\caption{Comparison of the energy, atomic forces, and stress tensor obtained by SPARC and  Quantum Espresso (QE). $f$ and $\sigma$ denote the maximum atomic force magnitude and the maximum stress tensor component, respectively.}
\centering
\begin{tabular}{c c c c c}
\toprule
System & & SPARC & QE & Difference \\
\midrule
\multirow{3}{*}{Br$_4$FeLi$_2$} & $E$~(ha/atom) & $-27.707195$ & $-27.707217$ & $2.14 \times 10^{-5}$\\
& $f$~(ha/bohr) & $\ 0.013182$ & $\ 0.013176$ & $5.87 \times 10^{-6}$ \\
& $\sigma$~(GPa) & $1.35$ & $1.43$ & $0.08$ \\
\midrule
\multirow{3}{*}{\centering\parbox{1cm}{FeLi$_4$N$_2$\newline(spin)}} & $E$~(ha/atom) & $-23.717767$ & $-23.717773$ & $5.48 \times 10^{-6}$ \\
& $f$~(ha/bohr) & $\ 0.244376$ & $\ 0.244382$ & $6.74 \times 10^{-6}$ \\
& $\sigma$~(GPa) & $21.11$ & $20.93$ & $0.18$ \\
\midrule
\multirow{3}{*}{MoO$_3$} & $E$~(ha/atom) & $-29.879153$ & $-29.879201$ & $4.80 \times 10^{-5}$ \\
& $f$~(ha/bohr) & $\ 0.012711$ & $\ 0.012609$ & $1.02 \times 10^{-4}$ \\
& $\sigma$~(GPa) & $4.82$ & $4.65$ & $0.17$ \\
\midrule
\multirow{3}{*}{TiO$_2$} & $E$~(ha/atom) & $-30.569899$ & $-30.569986$ & $8.77 \times 10^{-5}$ \\
& $f$~(ha/bohr) & $\ 0.011538$ & $\ 0.011455$ & $8.26 \times 10^{-5}$ \\
& $\sigma$~(GPa) & $16.17$ & $16.12$ & $0.05$ \\
\bottomrule
\end{tabular}
\label{table:QE_comp}
\end{table}

\subsection{Scaling and performance} \label{Subsec:Performance}
We next assess the scaling and performance of the real-space DFT+U framework. We begin by evaluating its strong scaling behavior in the context of an AIMD simulation, considering $2\times2\times2$, $3\times3\times3$, and $4\times4\times4$ supercells of rutile TiO$_2$, containing 48, 162, and 384 atoms and 384, 1296, and 3072 electrons, respectively. We consider isokinetic ensemble (NVK) simulations with a Gaussian thermostat \cite{minary2003algorithms} at a temperature of 300 K, while using a  time step of 2 fs. In each AIMD step, spin-unpolarized PBE+U calculations with $\Gamma$-point Brillouin zone integration are performed using $U = 5$~eV for the $d$-electrons in Ti. The grid spacing is chosen to be $0.3$ bohr and the radial orbital truncation threshold is chosen to be $5 \times 10^{-3}$ bohr$^{-3/2}$. These and other parameters are chosen such that the  energy is converged to within $\sim 10^{-3}$~ha/atom. We perform 10 AIMD steps to allow the per-step timings to stabilize before recording the timing data. The default parallelization in SPARC is used in all the calculations.

\begin{figure}[htbp]
    \centering
    \includegraphics[width=0.8\linewidth]{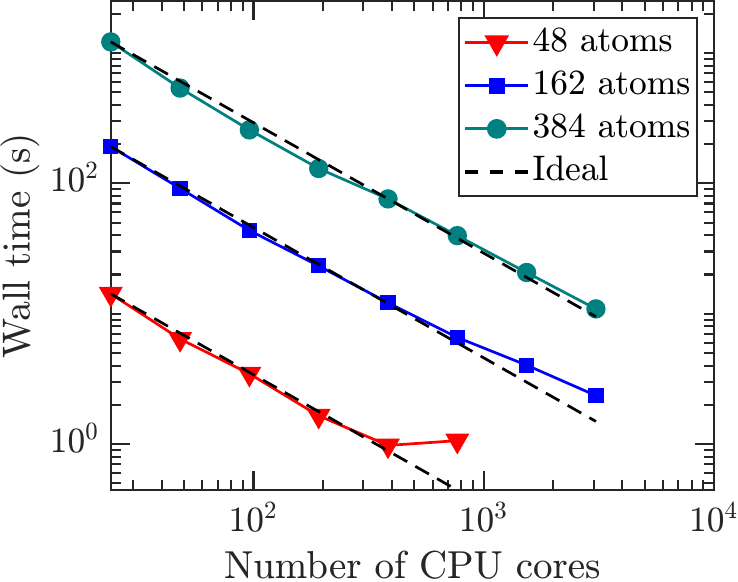}
    \caption{Strong scaling performance of an AIMD step in SPARC for TiO$_2$ systems with PBE+U. All parameters have been chosen so as to achieve accuracy of $10^{-3}$ ha/atom in the energy.}
    \label{fig:Scaling}
\end{figure}

In Fig.~\ref{fig:Scaling}, we present the strong scaling results obtained for the TiO$_2$ systems. Specifically, the total wall time per AIMD step, which includes three SCF iterations as well as the calculation of forces and the stress tensor, is plotted as a function of the number of CPU cores. The number of cores ranges from 24 to 3072 for the $(\text{TiO}_2)_{54}$ and $(\text{TiO}_2)_{128}$ systems, and from 24 to 768 for the $(\text{TiO}_2)_{16}$ system. We observe that the DFT+U framework in SPARC demonstrates excellent strong scaling up to thousands of cores, achieving $90.86$\%, $90.83$\%, and $87.18$\% efficiency for the $(\text{TiO}_2)_{16}$, $(\text{TiO}_2)_{54}$, and  $(\text{TiO}_2)_{128}$ systems on $384$, $768$, and $3072$ cores, respectively. As is to be expected, larger systems allow scaling to a larger number of cores. Due to the excellent scaling, the minimum time to solution for the $(\text{TiO}_2)_{16}$, $(\text{TiO}_2)_{54}$, and  $(\text{TiO}_2)_{128}$ systems is brought down to $0.98$, $2.36$, and $10.89$ s, respectively. Further reductions are indeed possible for the $(\text{TiO}_2)_{54}$ and  $(\text{TiO}_2)_{128}$ on using a larger number of cores. The results also indicate that the scaling with system size is approximately quadratic, with a scaling exponent of 2.1, consistent with typical scaling observed in DFT calculations for small to moderate sized systems.

In Table~\ref{table:QE_comp_scaling}, we compare the timings of SPARC and QE on the minimum, maximum, and  intermediate number of processors used in the strong scaling study. In particular, the intermediate number of processors is selected to yield a wall time close to the minimum for each code on the available resources. Due to the complexities associated with comparisons in an AIMD simulation, we restrict ourselves to the calculation of the electronic ground state, i.e., first AIMD step. In QE, we use a plane-wave cutoff of 33~ha and set the SCF tolerance suitably, ensuring that the energy is accurate to within $\sim 10^{-3}$~Ha/atom, commensurate with the accuracy in SPARC. Indeed, the discretizations and parameters in both SPARC and QE are chosen to achieve the target accuracy of $10^{-3}$~ha/atom, while maximizing computational efficiency. The results show that SPARC is competitive with QE even on the smallest core counts and becomes increasingly advantageous for larger systems. In particular, SPARC is $\sim 2.5 \times$ faster than QE for the $(\text{TiO}_2)_{128}$ system even on 24 cores. Unlike SPARC, QE is unable to efficiently utilize the largest core counts,  failing to execute for the $(\text{TiO}_2)_{54}$ and $(\text{TiO}_2)_{128}$ systems. In terms of minimum wall time, SPARC outperforms QE by more than an order of magnitude, delivering speedups of $\sim 8 \times$, $\sim 30 \times$, and $\sim 47 \times$ for the $(\text{TiO}_2)_{16}$, $(\text{TiO}_2)_{54}$, and $(\text{TiO}_2)_{128}$ systems, respectively.  These gains can be attributed to SPARC's superior parallel scalability and the locality of the atomic orbitals in real-space, which become increasingly beneficial as processor count and/or system size grows.

\begin{table}[htbp]
\caption{Comparison of the timings in seconds of SPARC and QE for calculation of the electronic ground state on the minimum (Min.), maximum (Max.), and  intermediate (Int.) number of processors used in the strong scaling study. The value in parentheses in the ``Int.'' column indicates the number of processors that yields the minimum wall time for each code on the available resources.``$-$'' indicates that the code failed to execute.}
\centering
\begin{tabular}{c c c c c}
\toprule
 Size & Code & Min.  & Max. & Int. \\
\midrule
\multirow{2}{*}{$(\text{TiO}_2)_{16}$} & SPARC & $14.20$ ($24$) & $1.06$ ($768$) & $0.98$ ($384$)\\
& QE & $13.43$ ($24$) & $12.87$ ($768$) & $7.73$ ($96$) \\
\midrule
\multirow{2}{*}{$(\text{TiO}_2)_{54}$}  & SPARC & $191.1$ ($24$) & $2.36$ ($3072$) & $2.36$ ($3072$) \\
 & QE & $175.0$ ($24$) & $-$ & $71.0$ ($384$) \\
\midrule
\multirow{2}{*}{$(\text{TiO}_2)_{128}$}  & SPARC & $1215$ ($24$) & $10.9$ ($3072$) & $10.9$ ($3072$) \\
& QE & $3020$ (24) & $-$ & $516$ ($384$) \\
\bottomrule
\end{tabular}
\label{table:QE_comp_scaling}
\end{table}

\subsection{Application: TiO$_2$ polymorphs} \label{Subsec:Polymorph}

The polymorphs of TiO$_2$ have attracted significant attention due to their relevance in a number of applications. In particular, their relative stability governs phase formation under varying thermodynamic conditions, while their distinct electronic structures and photocatalytic properties are important for technologies such as photovoltaics, photocatalysis, and solar energy conversion~\cite{curnan2015tio2,zhang2019tio2}. We now apply the real-space DFT+U framework to study the following TiO$_2$ polymorphs: rutile ($P4_2/mnm$), anatase ($I4_1/amd$), and brookite ($Pbca$). Specifically, we first examine how exchange-correlation consistency in the generation of local atomic orbitals influences the DFT+U results (Section~\ref{subSubsec:XCorb}), followed by a scheme for optimizing the Hubbard parameter (Section~\ref{subSubsec:U_opt}).

We consider 6-atom, 12-atom, and 24-atom cells for the rutile, anatase, and brookite phases of TiO$_2$, respectively. The structures correspond to the r$^2$SCAN-optimized geometries available through the Materials Project database \cite{materProj1,Horton2025}, and are held fixed during the simulations, i.e., the atom positions and cell dimensions are not relaxed. We perform spin-unpolarized calculations with Brillouin zone integration using a uniform wavevector grid with spacing of $\sim 0.11$ bohr$^{-1}$. We employ the r$^2$SCAN  exchange-correlation functional \cite{r2scan}, unless otherwise specified, and ONCV pseudopotentials from the SG-15 set \cite{sg15}, being free from NLCC. We employ 12-th order finite differences with a grid spacing of  $\sim 0.2$~bohr. This and other parameters are chosen to  ensure  that the energies are converged to within $10^{-4}$~ha/atom. 
  
\subsubsection{Exchange-correlation consistency for local atomic orbitals} \label{subSubsec:XCorb}

It is common for DFT+U calculations to employ atomic orbitals generated using local density approximation (LDA)/ generalized gradient approximation (GGA) exchange-correlation, even though  the calculations themselves utilize higher-rung exchange-correlation functionals such as r$^2$SCAN~\cite{r2scanu2025,zhang2019tio2,localProj_dftu}, which introduces an inconsistency. Leveraging the spectral single-atom radial solver~\cite{bhowmik2025} to generate exchange-correlation-consistent local orbitals, we investigate the relative thermodynamic stability of the rutile, anatase, and brookite phases of TiO$_2$ using r$^2$SCAN+U across a range of $U$ values, employing atomic orbitals generated with either r$^2$SCAN or PBE.

\begin{figure}[htbp]
    \centering
    \includegraphics[width=0.8\linewidth]{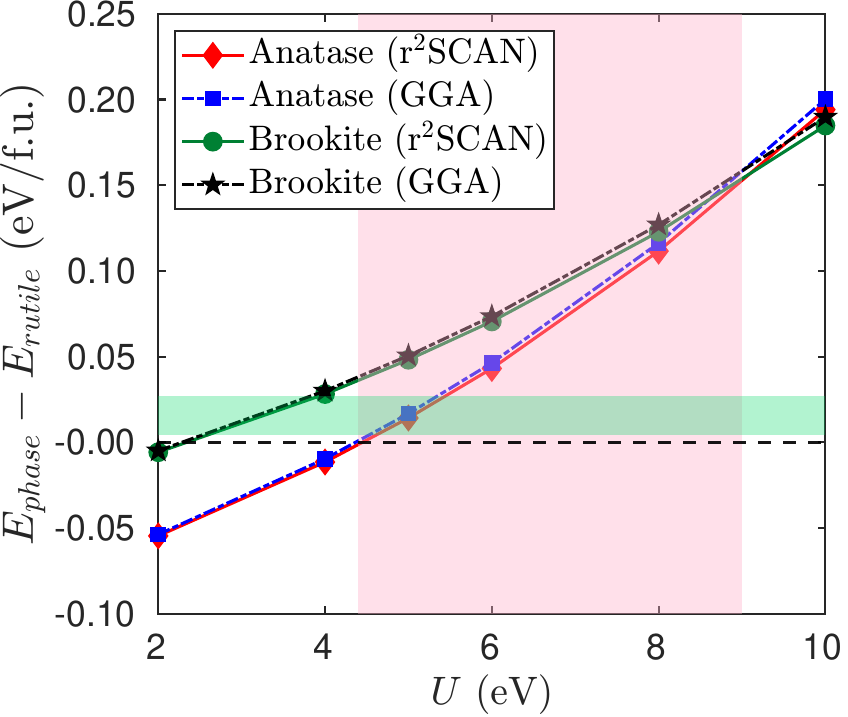}
    \caption{Variation in the DFT+U total energy differences between the anatase and brookite polymorphs of TiO$_2$ relative to the rutile polymorph, for atomic orbitals generated using either r$^2$SCAN or PBE. The green region marks the experimental $\left( E_{\text{anatase}} - E_{\text{rutile}} \right)$ range of values, while the pink region denotes the values of $U$ that result in the correct polymorph stability order.}
    \label{fig:xc_inconsistency}
\end{figure}

In Fig.~\ref{fig:xc_inconsistency}, we plot the energy differences of the anatase and brookite phases relative to the rutile phase as a function of $U$. Consistent with earlier studies \cite{curnan2015tio2}, values of $U$ between 4.4 and 9~eV maintains the correct ordering of the polymorphs: $E_{brookite} > E_{anatase} > E_{rutile}$, as observed in experiments \cite{Hanaor2011}. It is clear that the curves obtained with the r$^2$SCAN and PBE atomic orbitals are very similar. In particular, though the error incurred by the PBE atomic orbital increases with $U$ and can be significant, e.g., $0.0046$~ha/f.u. for the rutile phase at $U=10$ eV, in terms of energy differences between the different phases, the error incurred is relatively minor and is within $\sim 10^{-4}$~ha/f.u. We can therefore conclude that although the exchange-correlation inconsistency in the local orbitals can introduce significant error, these effects can  cancel out when considering differences, as evidenced by the relative energetics in the above TiO$_2$ polymorph stability study. However, for applications that target absolute energies, the use of exchange-correlation-consistent local orbitals can potentially be important.

\subsubsection{Hubbard parameter optimization using hybrid  functionals}\label{subSubsec:U_opt}

The results of DFT+U simulations can vary significantly depending on the value chosen for the Hubbard parameter $U$~\cite{zhang2019tio2, curnan2015tio2}. The value is typically selected to reproduce a physical property, such as the bandgap, in agreement with experimental measurements~\cite{saiG_scanu, xuCarter2019} or higher-level theoretical predictions~\cite{U_from_cRPA, U_from_cDFT}. Here, we investigate an alternative approach in which hybrid functionals, which combine a fraction of nonlocal Hartree-Fock exact exchange with local/semilocal exchange-correlation contributions, are used to optimize the value of $U$. In particular, for different values of the parameter $U$, the DFT+U electronic ground-state quantities are used to evaluate the  total energy corresponding to a hybrid functional. The value of $U$ that minimizes this non self-consistent energy is taken to be the optimized $U$, denoted henceforth as $U_{opt}$.

\begin{figure}[htbp]
    \centering
    \includegraphics[width=0.78\linewidth]{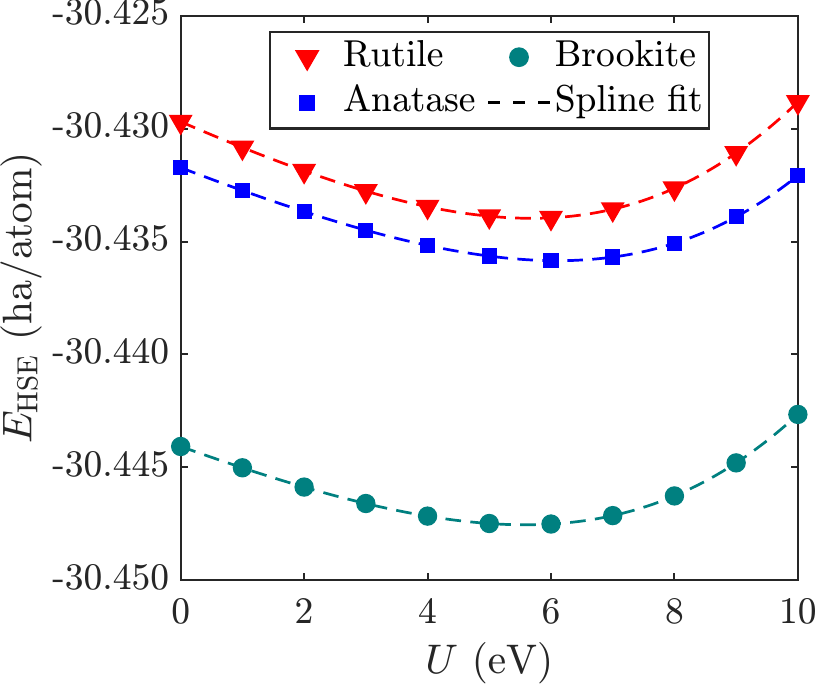}
    \caption{Non self-consistent HSE hybrid functional total energy evaluated at the r$^2$SCAN+U ground state as a function of the Hubbard parameter $U$.}
    \label{fig:U_opt}
\end{figure}

We have implemented this framework for the calculation of $U_{opt}$ in the M-SPARC electronic structure code. We consider the same three  polymorphs of TiO$_2$ as before: rutile, anatase, and brookite. We adopt the Heyd–Scuseria–Ernzerhof (HSE) \cite{hse} hybrid functional with 75\% exact exchange. The variation in the non self-consistent HSE total energy with $U$ for the different polymorphs is presented in Fig.~\ref{fig:U_opt}.  We observe a clear minimum in the energy for each phase, with the minima occurring at relatively similar $U$ values across the different phases. In particular, a  cubic spline fit to these data reveals minima at $U_{opt} = 5.66,\ 6.06,\ \text{and }5.56$~eV for the rutile, anatase, and brookite phases of TiO$_2$, respectively. 
 
\begin{figure*}[htbp]
  \centering

  \subfloat[\label{fig:Energy_conv}Rutile]{%
    \includegraphics[width=0.3\textwidth]{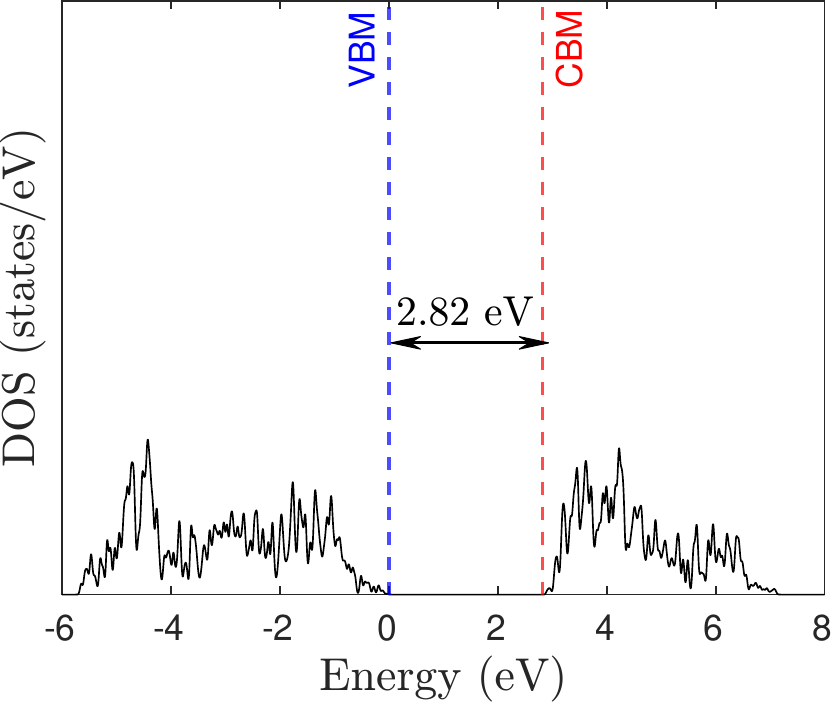}
  }
  \subfloat[\label{fig:Force_conv}Anatase]{%
    \includegraphics[width=0.3\textwidth]{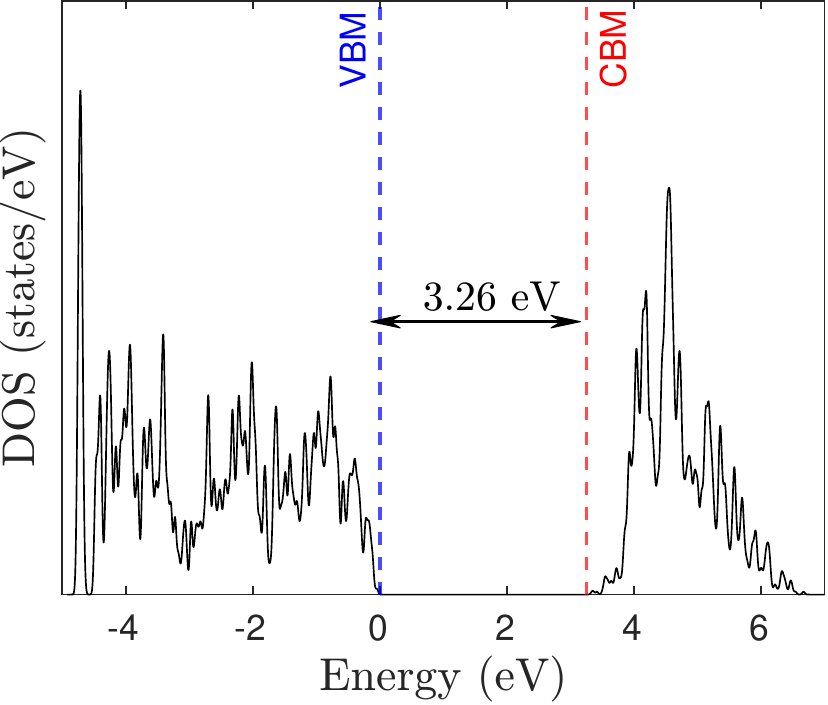}
  }
  \subfloat[\label{fig:Stress_conv}Brookite]{%
    \includegraphics[width=0.3\textwidth]{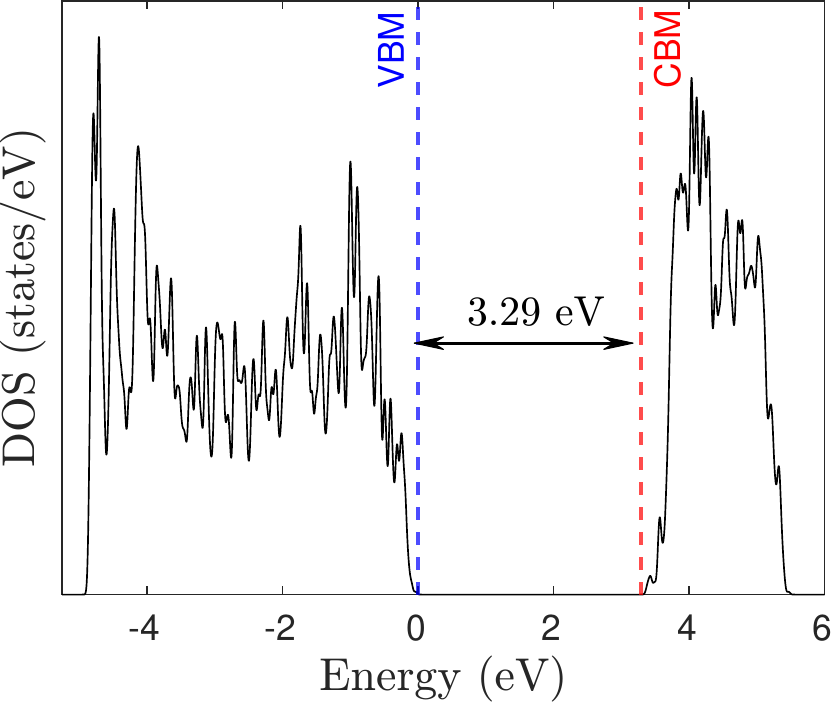}
  }
  \caption{DOS for the TiO$_2$ polymorphs using r$^2$SCAN+U, with optimized values $U_{opt}$.}
  \label{Fig:DOS}
\end{figure*}

We perform r$^2$SCAN+U calculations for the  TiO$_2$ polymorphs using their respective $U_{\text{opt}}$. In Fig.~\ref{Fig:DOS}, we plot the density of states (DOS) so obtained for the different phases. We find a band gap of 2.82, 3.26, and 3.29 eV for the rutile, anatase, and brookite phases, respectively, which are in very good agreement with the experimental values of 3.0, 3.2, 3.27   for the rutile \cite{rutileGapExpt}, anatase \cite{anataseGapExpt}, and brookite \cite{brookiteGapExpt} phases, respectively. Indeed, the  band gap values span a significant range for different choices of $U$. In particular, the bandgaps for the rutile, anatase, and brookite phases on choosing $U=\{0, 10\}$ eV  are $\{2.29, 3.04\}$, $\{2.55, 3.91 \}$, and $\{2.69, 3.86 \}$ eV, respectively, which makes the agreement with experiment  notable. The bandgap values obtained using HSE with 75\% exact exchange are nearly double those predicted here \cite{Badalov_2023}. Indeed, though $U_{opt}$ is determined from minimizing the hybrid energy, the band structure can still be noticeably different.  Previous studies using the linear response method have computed the $U$ values for the rutile phase to be in the range of $3.12$--$4.95$\,eV \cite{Xu2015, curnan2015tio2, u_first_princ_tio2}, at which the band gaps are significantly lower than experiment. Similar trends of underestimated band gaps\cite{u_first_princ_tio2} were observed for the anatase and brookite phases, for which the linear response $U$ values were calculated to be $2.92$\,eV and $2.93$\,eV, respectively.\cite{curnan2015tio2}

The energy of the rutile, anatase, and brookite phases at their respective $U_{opt}$ is $-29.883553$, $-29.877952$, and $-29.884041$ ha/atom, respectively. In terms of energy differences between the different TiO$_2$ phases at the $U_{opt}$ values, rutile is predicted to be more stable than anatase ($E_{anatase} - E_{rutile} = 0.46$~eV/f.u.), which is in agreement with experiment \cite{Hanaor2011}. In addition, the brookite phase is found to be slightly more stable than rutile ($E_{brookite} - E_{rutile} = -0.04$~eV/f.u.).  Though this result is  in contradiction with  experiments \cite{Hanaor2011}, it is in agreement with HSE calculations that employ 75\% exact exchange \cite{curnan2015tio2}, which was used to determine the $U_{opt}$ values, further demonstrating the effectiveness of the proposed approach. Note that the disagreement with experiments, despite the $U_{opt}$ values lying within the pink region of Fig.~\ref{fig:xc_inconsistency}, is due to the use of a common $U$ value for all phases in the results presented in that figure.

The proposed approach for optimizing the Hubbard parameter $U$ involves minimizing the hybrid functional energy.  However, this strategy does not necessarily yield correspondingly accurate geometric, electronic, or magnetic properties, as these depend on the derivatives of the energy. Moreover, the approach is presently limited to a single Hubbard parameter, i.e., $U_{Jl}=U$, and is therefore not directly applicable to systems where the Hubbard correction is applied to multiple atoms and/or to local atomic orbitals of different angular momenta, if distinct Hubbard parameter values are employed. Indeed, in such cases the formalism can be generalized to include constraints such as a target band gap, but this would introduce significant additional complexity.

Hybrid functionals contain an adjustable parameter that may themselves require tuning for the system under study. Recent efforts to address this limitation include matching experimental data while enforcing Koopmans' condition~\cite{Yang2023, Hyb_galli}, minimizing the random phase approximation (RPA) total energy~\cite{pitts2025selfconsistentrandomphaseapproximation}, and employing machine learning techniques to fit the parameter to data from higher-level theories~\cite{Liu2025, Khan_2025}. Hybrid functional calculations are also significantly more expensive than DFT+U calculations. Even though the proposed approach for the calculation of $U_{opt}$ does not require the self-consistent solution, it can still be associated with significant cost. To overcome this, the optimzation can be performed on a small representative system, e.g., unit cell. Note that the proposed scheme is in similar spirit to recent work that determines the optimized $U$ using diffusion Monte Carlo (DMC)~\cite{ghosh2024}. Hybrid functional calculations, however, are significantly less expensive than DMC and therefore more practical.

\section{Concluding remarks}\label{Sec:Concluding_remarks}
We developed an accurate and efficient framework for real-space DFT+U. Specifically, we obtained  expressions for the energy, atomic forces, and stress tensor amenable to real-space finite-difference discretization, and developed a large-scale parallel implementation in the SPARC electronic structure code. We verified the accuracy of the formalism through comparisons with established planewave results. We demonstrated that the implementation is highly efficient and scalable to thousands of processors, achieving over an order of magnitude reduction in minimum time to solution compared to established planewave codes, with increasing advantages as the system size and/or number of processors increased. We employed the real-space DFT+U framework to study the effects of exchange-correlation inconsistency in the generation of local atomic orbitals and introduced a scheme based on hybrid-functionals for optimizing the Hubbard parameter, both while studying  polymorphs of TiO$_2$.

Overall, the developed framework offers an attractive option for DFT+U calculations. Future research directions include the development of a GPU-accelerated implementation~\cite{sharma2023gpu}, as well as a highly efficient formalism for determining the Hubbard parameter, potentially leveraging machine learning techniques to enhance both accuracy and transferability. These advances have the potential to extend the applicability of DFT+U to a wider range of materials systems.


\section*{Supplementary Material}
SPARC and QE output files corresponding to the results presented in the manuscript. 

\section*{Acknowledgements}
The authors gratefully acknowledge the support of the U.S. Department of Energy, Office of Science, under Grant No. DE-SC0023445. This research was also supported by the supercomputing infrastructure provided by Partnership for an Advanced Computing Environment (PACE) through its Hive (U.S. National Science Foundation through Grant No. MRI1828187) and Phoenix clusters at Georgia Institute of Technology, Atlanta, Georgia.

\section*{Data Availability Statement}
The data that support the findings of this study are available within the article and the Supplementary Material.
\section*{Author Declarations}
The authors have no conflicts to disclose.

\section*{References}
\bibliography{Manuscript_Ref}

\end{document}